\documentclass[preprint,showpacs,preprintnumbers,amsmath,amssymb, superscriptaddress,longbibliography,nofootinbib]{revtex4-1}
\usepackage{graphicx} 
\usepackage{xcolor}
\usepackage{amsmath}
 \usepackage{hyperref}
\usepackage[top = 2cm, bottom = 2cm, right = 2cm, left =2cm]{geometry}
\usepackage{tikz}
\usepackage{caption}
\usepackage{cancel}
\usetikzlibrary{decorations.pathmorphing,calc}

\begin{document}

\title{Reissner Nordström black holes with integrable singularity interiors supported by string distributions}

\author{Milko Estrada }
\email{milko.estrada@gmail.com}
\affiliation{Departamento de Física, Facultad de Ciencias, Universidad de Tarapacá, Casilla 7-D, Arica, Chile}

\date{\today}

\begin{abstract}
The Reissner Nordström (RN) black hole is characterized by two well known pathologies: a central singularity and an inner horizon associated with potential instabilities and a loss of predictability. In this work, we show that the RN exterior geometry can arise from an interior spacetime containing an integrable singularity but no inner horizon. In this scenario, tidal forces remain finite near the origin, allowing nondestructive radial infall, while the conventional description in terms of a pointlike mass is replaced by an extended matter distribution. To illustrate this possibility, we provide explicit realizations of such an interior region based on string distributions, namely a cloud of strings (CS) and a newly defined fluid of strings (FS). While the standard cloud of strings model leads to a divergence in the conserved energy associated with timelike Killing vectors, the proposed FS model can be interpreted as a geometrically screened version of the string cloud distribution and admits configurations that, when extended to infinity, describe black holes with finite conserved energy. Physical consistency between the interior region and the RN exterior geometry requires the continuity of temperature across the interface, implying thermal equilibrium between the two regions, while discontinuities in the tangential pressure can signal gravitational phase transitions. These results determine the physical conditions under which string based interior distributions can consistently generate the RN exterior geometry and clarify the circumstances under which phase transitions at the event horizon may arise. 
\end{abstract}

\maketitle

\section{Introduction}

The observation of gravitational waves produced by the merger of two spinning black holes \cite{LIGOScientific:2016aoc} has firmly positioned black holes among the most compelling phenomena in gravitational physics. Black hole solutions are typically associated with the presence of a central singularity, where the geometric description of spacetime breaks down due to the divergence of curvature invariants. A widely studied approach to address this issue is provided by regular black holes (RBHs), in which the central singularity is usually replaced by a de Sitter core. However, these models generally possess an inner horizon, whose presence leads to well-known dynamical problems. In particular, the phenomenon of mass inflation can generate severe instabilities at this horizon \cite{Poisson:1990eh,Brown:2011tv}. More recent studies have reinforced this conclusion, indicating that instabilities associated with the inner horizon appear to be an unavoidable feature of astrophysically relevant RBHs \cite{Carballo-Rubio:2022pzu}. Moreover, it has been argued that the existence of such an inner horizon may compromise the predictability of spacetime beyond it \cite{Ovalle:2023vvu}.

An alternative route that has recently attracted attention consists of replacing the regular core with an integrable singularity (IS) \cite{Lukash:2013ts}. In this type of configuration, curvature invariants may diverge at the origin, while their volume integrals remain finite. An important consequence of this property is that tidal forces experienced by infalling observers can remain finite, suggesting that objects approaching the singularity radially may not necessarily be destroyed. Recent studies, such as \cite{Estrada:2023dcj}, have shown that these singularities can be weak according to Tipler’s classification \cite{Nolan:2000rn} and may allow the extension of radial geodesics through the origin . Furthermore, possible quantum interpretations of integrable singularities have also been explored \cite{Casadio:2023iqt}. Additional applications of this framework can be found in \cite{Ovalle:2023vvu,Estrada:2024moz,Arrechea:2025fkk}.

On the other hand, string theory suggests that the fundamental constituents of matter may be described as one-dimensional objects known as strings \cite{Younesizadeh:2022xzy}. In an effective gravitational framework, a macroscopic ensemble of such objects can be modeled as a cloud of strings, whose energy–momentum tensor modifies the surrounding spacetime geometry \cite{Letelier:1979ej}. This type of source has been widely used to study black hole models surrounded by string distributions \cite{Zafar:2025sxl,Javed:2026tdz,Gogoi:2025ied,Ahmed:2025ojg,Cai:2025pan}. A natural generalization of this model consists of considering a fluid of strings, in which the energy–momentum tensor includes an effective pressure or tension, adopting a structure similar to that of a perfect fluid \cite{Letelier:1983du}. Recently, this framework has been further generalized through the introduction of new equations of state for string fluids \cite{NunesdosSantos:2025alw}. However, when analyzing the conserved energy associated with timelike Killing vectors, it can be observed that the standard energy density of a cloud of strings leads to divergences in the conserved charge defined in \cite{Aoki:2020prb}. This feature suggests that the effective description of string distributions should be reconsidered when studying physically well-defined gravitational configurations. Motivated by this observation, in this work we introduce a modified version of the string-cloud energy density within the framework of a fluid of strings. This modification can be interpreted as a geometrical screening of the original distribution and leads to a finite conserved energy.

On the other hand, it is well known that the Schwarzschild black hole corresponds to a vacuum spacetime endowed with a central singularity. Traditionally, this solution is interpreted as being generated by a pointlike mass located at the origin. However, recent works have explored the possibility that the exterior Schwarzschild geometry may instead arise from more general internal configurations, obtained by matching an interior spacetime to the Schwarzschild exterior at the event horizon \cite{Ovalle:2024wtv,Maeda:2024tpl}. In this type of construction, the horizon can be interpreted as a surface separating two causally disconnected regions of spacetime, allowing the internal structure to be studied without modifying the geometry observed in the exterior region. This perspective naturally raises important physical questions regarding the nature of the matter capable of generating such an interior region. In particular, it is natural to ask whether similar configurations could also give rise to non-vacuum black hole geometries, such as electrically charged solutions, and whether an appropriate internal structure could simultaneously address the presence of a central singularity as well as the instabilities and loss of predictability typically associated with spacetimes containing an inner horizon.

In this context, it is natural to ask whether a similar configuration can be applied to charged solutions, such as the Reissner–Nordström (RN) black hole. This solution describes an electrically charged black hole and is characterized by the simultaneous presence of a central singularity and an inner horizon. The existence of this inner horizon introduces the well-known problems of instability and loss of predictability discussed above. Motivated by these considerations, in this work we investigate a scenario in which the exterior Reissner–Nordström geometry is generated by an extended interior region containing an integrable singularity and lacking an inner horizon.

To construct this scenario, we first analyze the general conditions that an interior geometry with an integrable singularity must satisfy, as well as the Israel–Darmois matching conditions between an interior region and an exterior black hole solution \cite{Israel:1966rt}. We further show that these junction conditions admit a natural thermodynamic interpretation: the temperature must remain continuous across the matching surface, while discontinuities in the tangential pressure can be interpreted as signals of phase transitions.

In order to investigate the nature of the matter capable of generating such an interior region for a Reissner–Nordström black hole, we study configurations in which the interior spacetime is described either by a cloud of strings or by the new fluid-of-strings model introduced in this work, while the exterior region corresponds to the Reissner–Nordström solution. In particular, we analyze the physical conditions that lead to the presence of an integrable singularity and the absence of an inner horizon. Finally, we determine the relations between the parameters of the interior string distributions and the exterior Reissner–Nordström geometry for which the matching conditions are satisfied and the temperature remains continuous at the interface. In addition, we identify the regions of the parameter space for which phase transitions at the event horizon may occur or be absent.

To clarify the structure of the paper, it is important to make explicit the connection between Section II and the subsequent sections. In Section II, we construct a new fluid-of-strings (FS) model defined over the entire radial domain $r \in [0,\infty)$, which exhibits an integrable singularity, absence of an inner horizon, and finite conserved energy. This construction should be understood as a baseline solution that is later employed, in Section VII, to model the interior region $r \in [0,h]$, where $h$ denotes the event horizon, of the exterior Reissner–Nordström (RN) geometry.

From Section III onward, we develop the theoretical framework required for this reinterpretation. In Section III, we define the geometric and matter setup, considering an interior region $r \in [0,h]$ in the presence of a negative cosmological constant, which, as we will see below, allows for the continuity of the radial pressure at the boundary $r=h$. In Section IV, we establish the conditions under which the interior region describes an integrable singularity, while in Section V we introduce the junction conditions. In Section VI, we analyze the case in which the interior region of the RN geometry corresponds to a cloud of strings, and in Section VII we study the case in which it is described by the fluid-of-strings model introduced in Section II.

\section{A new representation of a fluid of strings black hole model featuring an integrable singularity and finite conserved energy}
\label{SeccionFluido}

In this section, we first present a new string fluid model, showing that it can represent a black hole model over the entire domain $r \in [0,\infty)$. Below, in line with this work, we characterize it as the interior region of a Reissner–Nordström black hole. Reference \cite{Letelier:1983du} proposes a generalization of the cloud of strings model by incorporating pressure or tension into the energy–momentum tensor. This model is the so-called fluid of strings. See Appendix \ref{ApendiceFluidoDeCuerdas} for further details.  Due to the symmetries of the spacetime and in connection with the structure of the energy–momentum tensor \eqref{TensorEMfluid}, its general form is given by $T^t_t=T^r_r=-\rho^{(fs)} \,,\, T^\theta_\theta=T^\phi_\phi=p^{(fs)}$. It is of interest to establish an equation of state for the string fluid. In Ref. \cite{NunesdosSantos:2025alw}, an equation of state was proposed in which the energy density and the pressure are related through the form $\rho(r)^{(fs)}=\alpha(r)\cdot p(r)^{(fs)}$. In this reference, an equation of state with $\alpha(r)$ was proposed such that the model of Einstein’s equations represents a regular string-fluid geometry at the origin. Within this framework, the energy–momentum tensor takes the form
\begin{equation} \label{TensorEMFluidoCuerdas}
    (T^\mu_{\ \nu})^{(fs)}=\mathrm{diag}\left(-\rho(r)^{(fs)},-\rho(r)^{(fs)},\dfrac{\rho(r)^{(fs)}}{\alpha(r)},\dfrac{\rho(r)^{(fs)}}{\alpha(r)}\right) .
\end{equation}

We consider a line element 
\begin{equation} \label{ElementoDeLineaInterior}
ds^{2} = -f(r)_{(fs)}\,dt^{2} + \frac{dr^{2}}{f(r)_{(fs)}} + r^{2} d\Omega^{2}.
\end{equation}
As mentioned above, in this section the radial coordinate extends over the entire domain $r \in [0,\infty)$ and, consequently, there is an asymptotically flat boundary. As indicated in Ref. \cite{NunesdosSantos:2025alw}, within this framework the Einstein equations lead to a geometry of the form:
\begin{equation} \label{SolucionLuis1}
   f(r)_{(fs)}=1+\frac{c_2}{r}+\frac{c_1}{r}\,I(r),
   \end{equation}
\begin{equation} \label{SolucionLuis2}
    I(r)= \int dr \,\exp \left ( \frac{-2}{r \alpha(r)} \right).
\end{equation}
We are interested in a geometry that represents an integrable singularity near the origin. As we shall see below, a cloud of strings can represent such an integrable singularity. However, in this work we point out the following. The volume integral of the energy density can be interpreted as a global scalar functional that characterizes the spatial distribution of the material sector sourcing the geometry. In this regard, Ref. \cite{Aoki:2020prb} develops a formalism in which the integral $- \int_0^\infty 4\pi r^2 T^0_{\ 0}dr$ is associated with the energy as a conserved charge linked to a timelike Killing vector. Within this context, we note that for a cloud of strings the integral $\int_0^\infty 4\pi r^2 \rho \,dr \to \infty$, with $\rho$ given in Eq. \eqref{TensorEMNubeCuerdas1}, which becomes problematic for the definition of energy. Therefore, we shall test for the fluid of strings an energy density interpreted as a geometrical screening of the cloud of strings energy density. Subsequently, motivated by Ref. \cite{Aoki:2020prb}, we construct a new string fluid model with an equation of state $\rho(r)^{(fs)}=\alpha(r)\cdot p(r)^{(fs)}$. Thus, our energy density model is given by:
\begin{equation} \label{DensidadEnergiaDymnikovaLike}
    \rho^{(fs)} = \frac{M}{4\pi b^2 r^2} \exp(-r/b),
\end{equation}
where $b>0$ is a constant. Thus, the factor $\exp(-r/b)$ can be regarded as a form of geometrical screening of the cloud of strings energy density, i.e., $\rho_{cs} \sim \text{Const}/r^2 \to \rho^{(fs)} \sim (\text{Const}/r^2)\exp(-r/b)$. We can verify that the spatial integral of the energy density is finite
\begin{equation}
    \int_0^\infty 4\pi r^2 \rho^{(fs)} \,dr =M \Rightarrow \mbox{finite}\,.
\end{equation}

Thus, following the definition and assumptions of \cite{Aoki:2020prb}, our energy density, which screens the cloud-of-strings profile, yields a finite total energy, in contrast to the original cloud-of-strings profile. 

The $(t,t)$--$(r,r)$ and tangential components of the equations of motion are:
\begin{align}
- \frac{\left ( r (1-f_{(fs)})  \right)'}{r^2} =& {p}_r^{\,\,(fs)}=- {\rho}^{(fs)} \label{EqMovFluidoT} \,,\\
 \frac{\left ( r^2 (f_{(fs)})'\right)'}{2r^2} =& {p}_\theta^{\,\,(fs)}. \label{EqMovFluidoTan}
\end{align}

Thus, by evaluating our energy density profile, the temporal and radial components of the Einstein equations lead to the following solution 
\begin{equation} \label{SolucionDymnikovaLike}
    f(r)_{(fs)}=1-\frac{2M}{r} \left ( 1- \exp (-r/b)   \right) .
\end{equation}

By comparing Eqs. \eqref{SolucionLuis1} and \eqref{SolucionDymnikovaLike}, it is straightforward to note that $c_2=-2M$ and, moreover, that:
\begin{align}
    &2M\,\exp\!\left(-\frac{r}{b}\right)
=
c_1 \int dr\,\exp\!\left(-\frac{2}{r\,\alpha(r)}\right) \mbox{\,\,,we derive,} \nonumber \\
-&\frac{2M}{b}\exp (-r/b)
=
c_1\,\exp\!\left(-\frac{2}{r\,\alpha(r)}\right) \mbox{\,\,,we obtain:}\nonumber \\
& c_1=-\frac{2M}{b} \\
&\alpha(r)=\frac{2b}{r^2}
\end{align}

Thus, the components $-(T^0_{\ 0})^{(fs)}=-(T^1_{\ 1})^{(fs)}$ of the energy–momentum tensor \eqref{TensorEMFluidoCuerdas} are given by Eq. \eqref{DensidadEnergiaDymnikovaLike}, while the tangential components are given by:
\begin{equation} \label{TangencialDymnikovaLike}
    (T^3_3)^{(fs)}=(T^4_4)^{(fs)}=p_\theta^{\,\,(fs)}=\frac{M}{8\pi b^3} \exp(-r/b).
\end{equation}

\subsection{Integrable singularity}
The trace of the energy--momentum tensor \eqref{TensorEMFluidoCuerdas} is given by
$T^{(fs)} = -2\rho^{(fs)} + 2 p_\theta^{\,\,(fs)}=-R$, where $R$ is the Ricci scalar. Thus, we obtain the following relation:
\begin{align}
    & R=-T^{(fs)}= \frac{M}{2\pi b^2 r^2} \exp(-r/b)- \frac{M}{4\pi b^3}\exp(-r/b), \\
    &\Rightarrow R \big |_{r \to 0} \sim r^{-2} .\label{RicciFluidoOrigen}
\end{align}

Furthermore, replacing Eqs. \eqref{EqMovFluidoT} and \eqref{EqMovFluidoTan}, we obtain

\begin{equation} \label{EqMovimientoRicci}
  r^2 \cdot R=  2\left ( r (1-f_{(fs)})  \right)' - {\left ( r^2 (f_{(fs)})'  \right)'} .
\end{equation}

We note that although the Ricci scalar diverges near the origin, see Eq. \eqref{RicciFluidoOrigen}, the equations of motion \eqref{EqMovimientoRicci} are integrable and remain free of singularities. This situation can therefore be described as an integrable singularity.

\subsection{Absence of an inner horizon} 

\begin{figure}[ht]
  \begin{center}
      \includegraphics[width=3.5in]{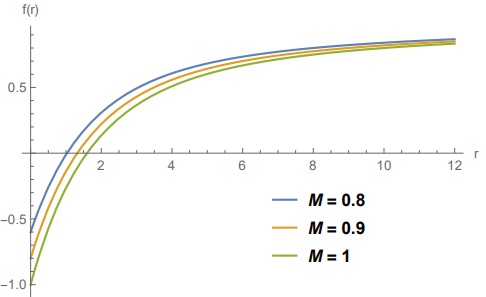}
         \caption{$f(r)_{(fs)}$ for $b=1,M=0.8,0.9,1$. We observe the absence of a potentially unstable inner horizon. }\label{FigFuncionFluido}
  \end{center}
\end{figure}

First, we note that function \eqref{SolucionDymnikovaLike} behaves near the origin as $f_{(fs)}|_{r \sim 0} \sim 1-2M/b$, with $2M/b>1$ in order for the metric signature to be $+,-,-,-$ near the origin. Moreover, $|1-\frac{2M}{b}|$ is sufficiently large and dominant in the vicinity of the origin, so as to avoid the appearance of a zero of the function $f(r)_{(fs)}$ near the origin. We observe the numerical behavior and the absence of a potentially unstable inner horizon in Fig. \ref{FigFuncionFluido} . Consequently, our model lacks the presence of an unstable de Sitter core, unlike other black hole solutions sourced by matter.

On the other hand, we note that the derivative $df_{(fs)}/dr = 2M/r^2 \left ( 1-\exp(-r/b)(1+r/b)  \right )$ is always positive. Consequently, since $f(0)_{(fs)}=f_{\min}<0$, there is only a single sign change, from negative to positive, at the event horizon. For both reasons discussed above, we can conclude that there is no presence of a potentially unstable inner horizon.

In this way, in this section we have presented a new string fluid model that represents a geometry with an integrable singularity and the absence of a potentially unstable inner horizon. As mentioned above, in this section we analyze a black hole model over the entire domain $r \in [0,\infty)$. Below, in line with this work, we characterize it as the interior region of a Reissner–Nordström black hole.

\section{Black Hole models with an Interior Geometry Featuring an Integrable Singularity} \label{SeccionConfiguracion}

We first analyze the physical configuration of generic matter models for both the interior and exterior regions. In the subsequent sections, we focus on the specific case in which the interior region is described by clouds and fluids of strings, while the exterior region is given by the Reissner–Nordström geometry. It is worth mentioning that the general framework presented in this section could also be of physical interest for future studies of other black hole exterior geometries, beyond RN, that possess a central singularity. Examples include black holes with cosmic void density profiles \cite{Lustosa:2025mxr}, black holes surrounded by dark matter \cite{Xu:2018wow}, hairy black holes \cite{Ovalle:2020kpd}, and quintessential black holes \cite{Kiselev:2002dx,Estrada:2025sku}.

In Fig. \ref{FigConfiguracion}, we schematically display the configuration of our geometry, which consists of two regions: an interior region, where the radial coordinate runs from the origin to the event horizon at $r = h$, i.e. $r \in [0,h]$, and an exterior region, extending from the event horizon to infinity (or to a cosmological horizon, depending on the case). Below, we describe both regions.

\begin{center}
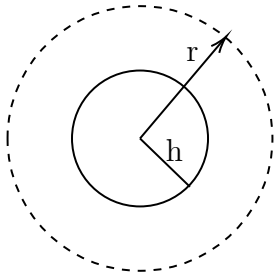

\tikzset{every picture/.style={line width=0.75pt}} 

\begin{tikzpicture}[scale=0.8,x=0.75pt,y=0.75pt,yscale=-1,xscale=1]

\draw[dashed] 
(205,101.5) .. controls (205,55.38) and (242.38,18) .. (288.5,18)
.. controls (334.62,18) and (372,55.38) .. (372,101.5)
.. controls (372,147.62) and (334.62,185) .. (288.5,185)
.. controls (242.38,185) and (205,147.62) .. (205,101.5) -- cycle ;

\draw
(245.63,101.5) .. controls (245.63,77.82) and (264.82,58.63) .. (288.5,58.63)
.. controls (312.18,58.63) and (331.38,77.82) .. (331.38,101.5)
.. controls (331.38,125.18) and (312.18,144.38) .. (288.5,144.38)
.. controls (264.82,144.38) and (245.63,125.18) .. (245.63,101.5) -- cycle ;

\draw (288.5,101.5) -- (320,132);

\draw (288.5,101.5) -- (341.71,38.53);
\draw [shift={(343,37)}, rotate = 130.2]
[line width=0.75] 
(10.93,-3.29) .. controls (6.95,-1.4) and (3.31,-0.3) .. (0,0)
.. controls (3.31,0.3) and (6.95,1.4) .. (10.93,3.29);

\draw (303,100) node [anchor=north west][inner sep=0.75pt] {h};
\draw (316,42) node [anchor=north west][inner sep=0.75pt] {r};

\end{tikzpicture}
\captionof{figure}{Schematic representation of the interior region $r \in [0,h]$ and the exterior region extending from $r = h$ to the black hole boundary.}
\label{FigConfiguracion}
\end{center}

\subsection{Description of the interior spacetime :}
It corresponds to a static and spherically symmetric spacetime, where the radial coordinate runs over $r \in [0,h]$.
The line element is given by:
\begin{equation} \label{ElementoDeLineaInterior}
ds^{2} = -f(r)\,dt^{2} + \frac{dr^{2}}{f(r)} + r^{2} d\Omega^{2},
\end{equation}
where $h$ corresponds to the event horizon, such that $f(h)=0$. Since this region represents the interior of the black hole, the signature of the metric tensor \eqref{ElementoDeLineaInterior} is $+,-,-,-$, and therefore $f(r)\le 0$ in this region. In this way, $f(r)$ is an increasing function, from its minimum value $f(r=0)=f_{\min}<0$ up to $f(h)=0$. First, we consider the presence of a cosmological constant in this region, where the matter sources in the interior are described by the interior energy–momentum tensor. 
\begin{equation} \label{TensorEMInterior}
(T^\mu_{\ \nu})^{(in)} = \mathrm{diag}\!\left(-\rho,\, p_r,\, p_\theta,\, p_\theta\right) . \end{equation}

The $(t,t)$--$(r,r)$ and tangential components of the equations of motion are:
\begin{align}
- \frac{\left ( r (1-f)  \right)'}{r^2}+ \Lambda =& -\rho =p_r \,, \\
 \frac{\left ( r^2 f'  \right)'}{2r^2}    + \Lambda =&p_\theta \,,
\end{align}
where we observe that $\rho = -p_r$. 

Thus, we can define the effective interior energy-momentum tensor 
$(\bar{T}^\mu_{\ \nu})^{(\text{in})}$, for $\Lambda=-3/l^2$, where $l$ denotes the AdS radius.
\begin{equation} \label{TensorEMInteriorEfectivo}
 (\bar{T}^\mu_{\ \nu})^{(\text{in})} = \mathrm{diag}\!\left(-\bar{\rho},\, \bar{p}_r,\, \bar{p}_\theta,\, \bar{p}_\theta\right)  .
\end{equation}
Thus, the  effective density and pressures are given by
\begin{align}
    \bar{\rho}=&\rho + \Lambda = \rho- \frac{3}{l^2}\,, \label{DensidadEfectiva} \\
    \bar{p}_r=& p_r- \Lambda = \frac{3}{l^2}+p_r=\frac{3}{l^2}-\rho=-\bar{\rho} \,,\label{PresionRefectiva} \\
    \bar{p}_\theta=& p_\theta- \Lambda=\frac{3}{l^2}+p_\theta \,. \label{presionTefectiva}
\end{align}

In this way, the equations of motion can be written as:
\begin{align}
- \frac{\left ( r (1-f)  \right)'}{r^2} =& \bar{p}_r=- \bar{\rho} \, \label{EqMovEfectivaInT}\\
 \frac{\left ( r^2 f'  \right)'}{2r^2} =& \bar{p}_\theta \,. \label{EqMovEfectivaInTan}
\end{align}

\subsection{ Description of the exterior spacetime} 

It corresponds to a static and spherically symmetric spacetime, where the radial coordinate runs over $r \in [h, r_b]$, with $r_b$ denoting the boundary of the spacetime. In the presence of a cosmological horizon, $r_b \to r_c$, where $r_c$ is the cosmological horizon. In the absence of the latter, i.e., in asymptotically flat or asymptotically AdS spacetimes, $r_b \to \infty$. The line element is given by:
\begin{equation} \label{ElementoDeLineaExterior}
ds^{2} = -f(r)_E\,dt^{2} + \frac{dr^{2}}{f(r)_E} + r^{2} d\Omega^{2}.
\end{equation}

The geometry is such that $f_E(h)=0$. Since this region represents the exterior of the black hole, the signature of the metric tensor~\eqref{ElementoDeLineaExterior} is $-,+,+,+$, and therefore $f_E(r)\ge0$ in this region. We consider the case without a cosmological constant in the external region, with an energy–momentum tensor of the form
\begin{equation} \label{TensorEMexterior}
    (T^\mu_{\ \nu})^{(E)} = \mathrm{diag}\left(-\rho^{(E)}, p_r^{(E)}, p_\theta^{(E)}, p_\theta^{(E)} \right) .
\end{equation}

It is straightforward to observe that the equations of motion can be obtained by replacing $f \to f_E$, $\bar{\rho} \to \rho^{(E)}$, $\bar{p}_r \to p_r^{(E)}$, and $\bar{p}_\theta \to p_\theta^{(E)}$ in Eqs. \eqref{EqMovEfectivaInT} and \eqref{EqMovEfectivaInTan}.

\section{ Representation of the INTERIOR REGION AS AN INTEGRABLE SINGULARITY} \label{SeccionInteriorSingularidadIntegrable}

In this section, we state the properties that a generic interior geometry must satisfy in order to represent an integrable singularity near the origin. In the following sections, we analyze the specific cases in which the interior regions correspond, respectively, to a string cloud and to the new string-fluid model provided in Section \ref{SeccionFluido}.

The trace of our efective interior energy--momentum tensor \eqref{TensorEMInteriorEfectivo} is given by  $\bar{T}^{(in)} = -2 \bar{\rho} + 2 \bar{p}_\theta$. Replacing Eqs. \eqref{EqMovEfectivaInT} and \eqref{EqMovEfectivaInTan}, we obtain
\begin{equation}
    -R =\bar{T}^{(in)} =- \frac{2\left ( r (1-f)  \right)'}{r^2}+ \frac{\left ( r^2 f'  \right)'}{2r^2} ,
\end{equation}
where $R$ is the Ricci scalar. Analogously to Section \ref{SeccionFluido}, this equation corresponds to the trace of the Einstein equations and can be rewritten as
\begin{equation} \label{EqMovimientoRicci1}
  r^2 \cdot R=  2\left ( r (1-f)  \right)' - {\left ( r^2 f'  \right)'} .
\end{equation}

Some remarks:

\begin{enumerate}
\item \label{Condicion1SI} We are interested in a function $f(r)$ that is finite at the origin and has no zeros in the interval $r \in [0,h]$. In this way, near the origin:
\begin{equation}
    f(r \sim 0)  \sim 1-a+ \mathcal{O}(r^{n})
\Rightarrow R \sim \frac{\mbox{Cte}}{r^2},
\end{equation}
where $a>1$ and $|1-a|$ is sufficiently large and dominant compared to the term $\mathcal{O}(r^{n})$ near the origin, in order to avoid a zero of the function $f(r)$ in the vicinity of the origin. As mentioned above, the absence of zeros in this interval implies the absence of a potentially unstable inner horizon. Although the Ricci scalar has a singularity $\sim r^{-2}$ at the origin, the equations of motion \eqref{EqMovimientoRicci1} are integrable, being free of singularities, and lead to a finite geometry $f(r)$ at the origin. This situation can be referred to as an integrable singularity. Under these assumptions, one can also observe that the Kretschmann scalar exhibits a singularity at $r = 0$.
\begin{equation}
    K = f''(r)^2 + \frac{4 f'(r)^2}{r^2} + \frac{4 (f(r) - 1)^2}{r^4}.
\end{equation}
\item \label{Condicion2SI} As mentioned, $f(r)$ is an increasing function from its minimum value $f(r=0)=f_{\min}<0$ up to $f(h)=0$. In this way, $df/dr>0$ for $r\in[0,h]$, with the temperature $T=\left.df/dr\right|_{h^-}=\left.df_E/dr\right|_{h^+}$. It is worth mentioning that this condition leads to the absence of an inner horizon.
\end{enumerate}

\section{Junction Conditions and Their Thermodynamic Implications} \label{SeccionMatching}

In this section, we first analyze the junction conditions between generic matter models describing both the interior and exterior regions. In the subsequent sections, we focus on the specific cases of interest in this work, in which the interior region is described by clouds and fluids of strings, while the exterior region is given by the Reissner–Nordström geometry. It is worth mentioning that the general junction conditions, together with their physical and thermodynamic implications presented in this section, could also be of interest for future studies of other black hole exterior geometries, beyond RN, that possess a central singularity, as in the examples mentioned in Section \ref{SeccionConfiguracion} \cite{Lustosa:2025mxr,Xu:2018wow,Ovalle:2020kpd,Kiselev:2002dx,Estrada:2025sku}. 

To construct a model describing a physically viable compact object, it is necessary to impose a smooth matching between the interior geometry and the exterior spacetime. For this purpose, we will consider the Israel–Darmois matching conditions \cite{Israel:1966rt}.

\begin{enumerate}
\item \label{Condicion1JC} At the surface $\Sigma$, identified with the event horizon radius $r = h$, these conditions imply the continuity of the geometry, known as the first fundamental form
\begin{equation} \label{PrimeraForma}
    [ds^2]_{\Sigma  (r=h)} = 0 \Rightarrow f(h)=f_E(h)=0,
\end{equation}
where $[F]_{\Sigma \,\equiv \,r=h} \equiv F(h^{+}) - F(h^{-})$.

\item \label{Condicion2JC} In this case,the second fundamental form is given by $[G_{\mu\nu} x^{\nu}]_{\Sigma(r=h)} = 0$, where $x^{\nu}$ is a unit vector projected along the radial direction. This implies the following
\begin{align} \label{SegundaForma}
    \bar{p}_r (h)=&p_r^{(E)} \,, \nonumber \\
    \frac{1-h f'(h)}{h^2}=&\frac{1-h \cdot \big (f_E(h) \big )'}{h^2} \,, \nonumber \\
                         \Rightarrow& f'(h)=\big (f_E(h) \big )' \,, \nonumber \\
                         \Rightarrow& T_{in}=T_{E}=T .
\end{align}

Since the derivative of $f'(h)$ matches the derivative of the exterior region at the event horizon $\big (f_E(h) \big )'=T$, in this work we also refer to the derivative of $f'(h)$ as the temperature.
Thus, the second fundamental form of the Israel–Darmois junction conditions leads to the continuity of the temperature between the interior and exterior regions.

\item \label{Condicion3JC} As we will see below, this condition is not mandatory, unlike the two previous ones. More specifically, this condition represents a consequence of the discontinuity (or continuity) of the tangential pressure in black hole thermodynamics. We obtain the interior temperature from Eq. \eqref{EqMovEfectivaInTan}. In an analogous way, we obtain the exterior temperature for the corresponding region. Thus, we evaluate Eq. \eqref{SegundaForma}

\begin{align}
    T_{in} =&   T_E  \,,\nonumber \\
 \cancel{h}\left(\bar p_\theta(h) - \frac{f''(h)  }{2}\right) =& \cancel{h} \left( p_\theta^{(E)}(h) - \frac{\big (f''(h) \big )_E }{2} \right) ,\nonumber \\
     \frac{3}{l^2}+p_\theta (h) - \frac{f''(h)  }{2} =& p_\theta^{(E)}(h) - \frac{\big (f''(h) \big )_E }{2} . \label{IgualdadTemperaturas3forma}
\end{align}
We use 
\begin{align}
    C_{in} = T_{in} \frac{dS}{dh} \left ( \frac{dT}{dh} \right )^{-1}  = T_{in} \frac{dS}{dh} \left ( \frac{d}{dh} \, \left ( \frac{df}{dh} \right) \right )^{-1} &\Rightarrow f''(h)= \frac{T}{C_{in}} \frac{dS}{dh} \,, \\
    \mbox{analogously} \,\, & \Rightarrow \big (f''(h) \big )_E = \frac{T}{C_{E}} \frac{dS}{dh} .
\end{align}

We assume that the entropy follows the area law, i.e. $S= \pi h^2$, thus evaluating \eqref{IgualdadTemperaturas3forma}
    
     \begin{align} \label{TransicionFase}
     \frac{C_{\text{in}} - C_E}{C_{\text{in}} \cdot C_E }
=& \frac{1}{T \pi h}\, \left (p_\theta^{(E)}(h) - \left( \frac{3}{l^2} + p_\theta(h)    \right ) \right ) ,\nonumber \\
\Rightarrow C_{\text{in}} - C_E
\sim & \,\,\,  p_\theta^{(E)}(h)-\bar{p}_\theta(h),
\end{align}
thus
\begin{align}
    \mbox{if} \,\, \bar{p}_\theta(h) = p_\theta^{(E)}(h) \Rightarrow & C_{\text{in}} = C_E \,,\\
    \mbox{if} \,\, \bar{p}_\theta(h) \neq p_\theta^{(E)}(h) \Rightarrow & C_{\text{in}} \neq C_E \,.
\end{align}

 In our case the entropy is continuous. In this way, we test that the first derivative of the Gibbs potential, $S = -dG/dT$, is also continuous. On the other hand, using $C = T\,dS/dT = -\,T\,d^2 G/dT^2$, we observe that {\it a discontinuity in the tangential pressure implies the existence of a second-order phase transition}.
\end{enumerate}

\section{Scenario 1: A Reissner–Nordström Exterior with a Cloud of Strings as the Interior Region with an Integrable Singularity} \label{SeccionRNconNube}

In Section ~\ref{SeccionConfiguracion}, we described our geometric setup to represent both the interior and exterior regions of a generic black hole geometry. In Section~\ref{SeccionInteriorSingularidadIntegrable}, we formulated the conditions that an interior region must satisfy in order to represent an integrable singularity near the origin, while in Section~\ref{SeccionMatching} we presented the junction conditions between the interior and exterior regions. However, the criteria introduced above must be tested in explicit black hole geometries. In this spirit, motivated by the existence of an inner horizon and a non-integrable singularity in the Reissner–Nordström solution (RN), we analyze two illustrative examples. In the present section, we study a Reissner–Nordström exterior with a cloud of strings as the interior region. In the following section, we consider a Reissner–Nordström exterior with the new string-fluid model provided in Section~ \ref{SeccionFluido}.

\subsection{Remarks on the Reissner–Nordström (RN) Black Hole} 

The electromagnetic field has an energy-momentum (EM) tensor that is symmetric and traceless, constructed from the Maxwell field tensor $F_{\mu\nu} = \nabla_\mu A_\nu - \nabla_\nu A_\mu$. The Maxwell equations are given by $\nabla_\mu F^{\mu\nu} = 0$ and $\nabla_{[\gamma} F_{\mu\nu]} = 0$. Solving the Einstein–Maxwell equations for a static, spherically symmetric electric charge leads to the RN metric. In this case, the metric function $f_E(r)$ appearing in the line element~\eqref{ElementoDeLineaExterior}, expressed in natural units, is given by
\begin{equation} \label{MetricaRN}
    f(r)_E = 1 - \frac{2M}{r} + \frac{Q^2}{r^2},
\end{equation}
where $M,Q$ correspond to the mass and electric charge, respectively. The inner and event horizons, $r_-$ and $h$, are given, respectively, by:
\begin{align} 
    &r_- = M - \sqrt{M^2-Q^2} \,,\\
    &h = M + \sqrt{M^2-Q^2} \,. \label{horizonteRN}
\end{align}

Thus, we note that it must be satisfied that $M > Q$ for $M,Q>0$, $\Rightarrow M^2-Q^2>0$. As we have mentioned, in our setup, the exterior geometry extends over $r \in [h, r_b]$, with the boundary corresponding to the radial coordinate approaching infinity, since the solution is asymptotically flat. Therefore, we are interested in the aforementioned range of the radial coordinate. The temperature is given by:
\begin{equation}
  T= \frac{1}{4\pi}  \left. \frac{df_E}{dr} \right|_{r=h} = \frac{1}{4\pi} \frac{2 \sqrt{M^2 - Q^2}}{\left(M + \sqrt{M^2 - Q^2}\right)^2} \,.
\end{equation}

The exterior energy–momentum tensor \eqref{TensorEMexterior} takes the form:
\begin{equation} 
    (T^\mu_{\ \nu})^{(E)} = \mathrm{diag}\left(-\rho^{(E)}=-E^2(r), p_r^{(E)}=-E^2(r), p_\theta^{(E)}=E^2(r), p_\theta^{(E)}=E^2(r) \right).
\end{equation}

For our line element, the electric field is given by
\begin{equation}
E(r) = \sqrt{-F_{01} F^{01}} = \frac{Q}{r^2} \, .
\end{equation}

\subsection{Representation of a Cloud of Strings as an Interior Region with an Integrable Singularity}

In this subsection, we explore a possible nature of the matter sources that could give rise to an interior region characterized by an integrable singularity. Specifically, we consider a cloud of strings as a potential matter source. In this context, as previously mentioned, the embedding of an ensemble of 1-branes—the so-called cloud of strings—can deform the spacetime geometry \cite{Letelier:1979ej}. See Appendix \ref{ApendiceNubeDeCuerdas} for a more detailed description of a cloud of strings.

The interior energy–momentum tensor $(T^\mu_{\ \nu})^{(in)}$, equations \eqref{TensorEMInterior}, reported in reference \cite{Letelier:1979ej}, Eq.~\eqref{TensorEMNubeCuerdas}, has the form 
\begin{equation} \label{TensorCloudAbril1}
    (T^\mu_{\ \nu})^{(in)}=\mbox{diag} \left ( -\rho, -\rho,0,0 \right ) \,,
\end{equation}
where the energy density originates from Eq.~\eqref{TensorEMNubeCuerdas1} has the form
\begin{equation} \label{TensorCloudAbril2}
    \rho=\frac{a}{r^2} .
\end{equation}

It is worth mentioning that, as discussed in Refs. \cite{Letelier:1983du,NunesdosSantos:2025alw}, a natural generalization of a cloud of strings consists in considering a fluid of strings, in which tangential pressure components are included in the energy–momentum tensor (see Eq. \eqref{TensorEMfluid} in Appendix \ref{ApendiceFluidoDeCuerdas}), adopting a structure analogous to that of a perfect fluid. In the limit where the tangential pressures vanish, the components of the energy–momentum tensor of a string fluid reduce to those of a cloud of strings, as given in Eq2. \eqref{TensorCloudAbril1} and \eqref{TensorCloudAbril2}.

Thus, from equations \eqref{DensidadEfectiva},\eqref{PresionRefectiva}, \eqref{presionTefectiva}, we note that the interior effective energy density and effective pressures take the following form
\begin{equation} \label{DensidadInteriorEfectivaNube}
    \bar{\rho}=-\bar{p}_r=\frac{a}{r^2}- \frac{3}{l^2} \,\,,\,\, \bar{p}_\theta=\frac{3}{l^2}\,.
\end{equation}

Thus, we can observe that the trace of the effective interior energy-momentum tensor $(\bar{T}^\mu_{\ \nu})^{(\text{in})}$, given by equations \eqref{TensorEMInteriorEfectivo}, \eqref{DensidadEfectiva}, \eqref{PresionRefectiva} and \eqref{presionTefectiva} is
\begin{equation}
  \bar{T}^{(\text{in})} = -2 \bar{\rho} + 2 \bar{p}_\theta = -\frac{2a}{r^2} + \frac{12}{l^2} = -R.  
\end{equation}

Thus, we note that the energy-momentum tensor behaves as $R \sim r^{-2}$ at short length scales. This implies that the equations of motion \eqref{EqMovimientoRicci1} are integrable near the origin, being free of singularities. Thus, condition \ref{SeccionInteriorSingularidadIntegrable} \ref{Condicion1SI} is satisfied for the string cloud model to represent the interior region as an integrable singularity.

Below, we present a solution to the equations of motion characterized by the sources of a cloud of strings together with a cosmological constant, as given in Eq. \eqref{DensidadInteriorEfectivaNube}. The solution of the equations of motion~\eqref{EqMovEfectivaInT} and~\eqref{EqMovEfectivaInTan}, together with the trace equation~\eqref{EqMovimientoRicci}, gives rise to the following metric function $f(r)$, which remains finite at the origin
\begin{equation} \label{metricaCS}
    f(r)=1-a + \frac{r^2}{l^2} \,,
\end{equation}
where $a>1$ and $|1-a|$ is sufficiently large compared to the term $r^2/l^2$ near the origin, in order to avoid a zero of the function $f(r)$ in the vicinity of the origin. We have set the integration constants to zero since they lead to a singularity in the metric tensor. This result is further reinforced by the fact that, in this work, we are interested in an interior geometry associated with an integrable singularity, for which the Ricci tensor is such that the equations of motion remain integrable.

In order to test conditions \ref{SeccionInteriorSingularidadIntegrable} and \ref{Condicion2SI} for the interior region to be represented as an integrable singularity without an inner horizon, we note that $f(r)$ is an increasing function, starting from its minimum value $f(r=0)=1-a<0$, with $a>1$, and reaching $f(h)=0$. Indeed, $df/dr = 2r/l^{2} > 0$ for $r \in [0,h]$, and therefore the function is monotonically increasing and negative throughout this interval. Therefore, we point out that the parameter $a$ must satisfy $a>1$ in order for this condition to be fulfilled.

\subsection{Junction conditions:}

As mentioned in Equation \eqref{horizonteRN}, the event horizon of the external RN region is given by $h = M + \sqrt{M^2-Q^2}$.
\begin{itemize}
\item In order to satisfy junction condition \ref{SeccionMatching}.\ref{Condicion1JC}, i.e. the so-called first fundamental form of the Israel–Darmois formalism, upon evaluating the functions \eqref{MetricaRN} and \eqref{metricaCS} on \eqref{PrimeraForma}, one obtains the following condition
\begin{equation}
    a= 1 + \left ( \frac{M + \sqrt{M^2-Q^2}}{l} \right )^2 \,,
\end{equation}
which is consistent with the previously mentioned condition that $a > 1$.
\item In order to satisfy the junction condition \ref{SeccionMatching}.\ref{Condicion2JC}, i.e., the so-called second fundamental form of Israel–Darmois, and by using Equation \eqref{SegundaForma}, the continuity of the temperature at the event horizon $r = h$ leads to the following value of the cosmological constant in the interior region:
\begin{equation}
 \Lambda=  - \frac{3}{l^2} = - \frac{3 \sqrt{M^2 - Q^2}}{\left(M + \sqrt{M^2 - Q^2}\right)^3}.
\end{equation}

It is worth noting that, in Section \ref{seccionElectrico} below, we discuss the discontinuity of the electric field between the interior and exterior regions.

\item As mentioned, the third condition III.\ref{Condicion3JC} is not mandatory like the previous two, since it does not stem from the Israel–Darmois junction conditions. However, it is useful to test whether a phase transition occurs at the event horizon $r = h$. Using Equation \eqref{TransicionFase}:
\begin{equation}
    C_{\text{in}} - C_E
\sim  \bar{p}_\theta(h) - p_\theta^{(E)}(h)  = -\Lambda - \frac{Q^2}{h^4} = 
\frac{3 M \left(M + \sqrt{M^2 - Q^2}\right) - 4 Q^2}{\left(M + \sqrt{M^2 - Q^2}\right)^4}.
\end{equation}

Testing the numerator of the last equation for zero leads us to:
\begin{align}
  Q^2 \left( \frac{16 Q^2}{9 M^2} - \frac{5}{3} \right) = 0 \Rightarrow Q_c^2= \frac{15}{16} M^2 \,,
\end{align}
where the previously mentioned condition $M^2 > Q^2$ is satisfied for $Q=Q_c$. We have discarded $Q = 0$, since this would lead to an external solution without electric charge. That is, for all values of $Q^2 \neq Q_c^2$, a second-order phase transition occurs at $r = h$ for the parameters mentioned above. Conversely, for $Q^2 = Q_c^2$, no phase transition occurs at this location.
\end{itemize}

\section{Scenario 2: A Reissner–Nordström Exterior with a new fluid of Strings as the Interior Region with an Integrable Singularity} \label{SeccionRNconFluido}

In Section \ref{SeccionFluido}, a new string fluid model was introduced. There, it was presented under the assumption that it extends over the entire spacetime, $r \in [0,\infty)$. In the present section, we analyze instead the case in which the model of Section \ref{SeccionFluido}, in the presence of a negative cosmological constant, is defined only in the interval $r \in [0,h]$, i.e., it can represent the interior region of the Reissner–Nordström solution, as depicted in Scheme \ref{FigConfiguracion}. It is worth mentioning that, in Section \ref{SeccionRNconNube}, we have already highlighted the main features of the Reissner–Nordström solution that are relevant to this work.

\subsection{ Representation of the New Fluid of Strings model as an Interior Region with an Integrable Singularity}

The effective interior energy–momentum tensor \eqref{TensorEMInteriorEfectivo}, obtained from Eqs.~\eqref{DensidadEfectiva}, \eqref{PresionRefectiva}, and \eqref{presionTefectiva}, \eqref{DensidadEnergiaDymnikovaLike} and \eqref{TangencialDymnikovaLike}, takes the following form

\begin{align}
    &\bar{\rho}=-\bar{p}_r=\frac{M}{4\pi b^2 r^2} \exp(-r/b)- \frac{3}{l^2}, \\ &\bar{p}_\theta=\frac{M}{8\pi b^3} \exp(-r/b)+\frac{3}{l^2}.
\end{align}

We can observe that the trace of the effective interior energy–momentum tensor $(\bar{T}^\mu_{\ \nu})^{(\text{in})}$, which includes the contribution of the cosmological constant, is given by
\begin{equation}
  \bar{T}^{(\text{in})} = -2 \bar{\rho} + 2 \bar{p}_\theta = -\frac{M}{2\pi b^2 r^2} \exp(-r/b)+ \frac{M}{4\pi b^3}\exp(-r/b) + \frac{12}{l^2} = -R.  
\end{equation}

Thus, the trace of the effective interior energy–momentum tensor with a cosmological constant behaves as $R \sim r^{-2}$ at short length scales. This behavior implies that the equations of motion \eqref{EqMovimientoRicci} are integrable in the vicinity of the origin and remain free of pathological singularities. Consequently, condition \ref{SeccionInteriorSingularidadIntegrable} \ref{Condicion1SI} is satisfied, allowing this new string fluid model to consistently represent the interior region as an integrable singularity. The geometry is given by:

\begin{equation} \label{SolucionDymnikovaLikeConstanteCosmologica}
    f(r)=1-\frac{2M}{r} \left ( 1- \exp (-r/b)   \right) +  \frac{r^2}{l^2} \,,
\end{equation}
which behaves as $f\mid_{r\sim 0}\sim 1-\frac{2M}{b}+\frac{r^2}{l^2}$ near the origin, with $\frac{2M}{b}>1$ and $|1-\frac{2M}{b}|$ sufficiently large compared to the term $\frac{r^2}{l^2}$ in the vicinity of the origin, so as to avoid a zero of the function $f(r)$ near the origin.

In order to test condition \ref{SeccionInteriorSingularidadIntegrable} \ref{Condicion2SI} for the interior region to be represented as an integrable singularity without an inner horizon, we note that $f(r)$ is an increasing function, starting from its minimum value $f(r=0)=1-\frac{2M}{b}<0$, with $\frac{2M}{b}>1$, and, as we shall see below, reaching $f(h)=0$. Indeed, $df/dr=\frac{2M}{r^2}\left(1-\exp(-r/b)\left(1+\frac{r}{b}\right)\right)+\frac{2r}{l^2}>0$ for $r\in[0,h]$.

\subsection{Junction conditions:}

In order to satisfy junction condition \ref{SeccionMatching} \ref{Condicion1JC}, i.e. the so-called first fundamental form of the Israel–Darmois formalism, upon evaluating the functions \eqref{MetricaRN} and \eqref{SolucionDymnikovaLikeConstanteCosmologica} on \eqref{PrimeraForma}, one obtains the following conditions:
\begin{equation} \label{Israel1Fluido}
    \frac{2M}{h} \exp (-h/b)= \frac{Q^2l^2-h^4}{h^2l^2}\,,
\end{equation}
where the event horizon $h$ is given by Eq. \eqref{horizonteRN}. The previous equation, in turn, leads to the following:
\begin{equation} \label{CondicionDeL}
    l^2> \frac{h^4}{Q^2}.
\end{equation}

In order to satisfy the junction condition \ref{SeccionMatching} \ref{Condicion2JC}, i.e., the so-called second fundamental form of Israel–Darmois, and by using Equation \eqref{SegundaForma}, the continuity of the temperature at the event horizon $r = h$ leads to the following condition:
\begin{equation} \label{Israel2Fluido}
    \frac{2M}{h} \exp (-h/b)=\frac{2b(Q^2l^2+h^4)}{h^2l^2(b+h)} .
\end{equation}

We can observe that the left-hand sides of Eqs. \eqref{Israel1Fluido} and \eqref{Israel2Fluido} coincide. By equating both equations, we obtain the following value for the AdS radius
\begin{equation} \label{ValorDeL}
   l^2= \frac{h^4}{Q^2} \cdot \left ( \frac{3b+h}{h-b} \right ) .
\end{equation}

Thus, for this value of $l^2$, with $b,h>0$ we must evaluate condition \eqref{CondicionDeL}
\begin{align}
    \frac{3b+h}{h-b}>1 &\Rightarrow \frac{4b}{h-b}>0 \,, \nonumber \\
                       &\Rightarrow h>b \,, \label{Condicionh>b}
\end{align}
where $h$ is given by Eq. \eqref{horizonteRN}. Thus, condition \eqref{Condicionh>b} must be satisfied. It is worth noting that, in Section \ref{seccionElectrico} below, we discuss the discontinuity of the electric field between the interior and exterior regions.

As mentioned, the third condition \ref{SeccionMatching} \ref{Condicion3JC} is not mandatory like the previous two, since it does not stem from the Israel–Darmois junction conditions. However, it is useful to test whether a phase transition occurs at the event horizon $r = h$. Using Equation \eqref{TransicionFase}:

\begin{align}
    C_{\text{in}} - C_E
\sim  \bar{p}_\theta(h) - p_\theta^{(E)}(h)  = \frac{h}{16\pi b^3} \frac{2M}{h} \exp(-r/b)+\frac{3}{l^2}  - \frac{Q^2}{h^4}   \,,                 
\end{align}
using equations \eqref{Israel1Fluido} and \eqref{ValorDeL}
\begin{equation}
C_{\text{in}} - C_E
\sim  \bar{p}_\theta(h) - p_\theta^{(E)}(h)    = - \frac{Q^2 (24\pi b^3-8\pi b^2 h - h^3)}{4\pi b^2h^4 (3b+h)} .
\end{equation}

Numerically, we find that the last equation vanishes for
\begin{equation}
    b=b_c =0.4116145346 \cdot h \,.
\end{equation}

Hence, there is no phase transition at $b = b_c$, whereas a phase transition occurs for $b \neq b_c$.

\section{A brief remark on the electric field.} \label{seccionElectrico}

From a physical perspective, the geometrical structure of the space--time in each region has different origins: in the interior it is governed by a cloud of strings, whereas in the exterior the electromagnetic tensor is responsible for generating a Reissner–Nordström geometry. Thus, the electric fields is such that \begin{equation}
    F_{tr}^{\text{(out)}} = \frac{Q}{r^2}, \qquad F_{tr}^{\text{(in)}} = 0.
\end{equation}
Usually, the presence of an electric field outside a hypersurface, which simultaneously vanishes in its interior, is treated by assuming a thin shell endowed with its own surface energy–momentum tensor \cite{Lemos:2021jtm,Abellan:2025bde}. In contrast, in our case the geometrical transition between the interior and exterior regions is smooth, in the sense that both the metric tensor and the radial derivative of the function $f(r)$ are continuous, and the Israel conditions lead to the continuity of the radial pressure in the vicinity of the event horizon. In this way, we examine whether the limiting behavior of the electromagnetic field near the horizon allows one to define an effective surface charge density in the exterior vicinity of the event horizon. We define the projected electromagnetic field \cite{Lemos:2021jtm}:
\begin{equation} \label{FaradayProyectado}
F_a^{\,(+)} = \displaystyle \lim_{r \to h^+} F_{\alpha\beta}^{\text{(out)}}\, e^\alpha_a \, n^\beta_{\,(+)}
\end{equation}

As we will see below, in the present context we are interested in evaluating the jump of the projected electromagnetic field in the limits $r \to h^\pm$. To this end, the projected electromagnetic field can be consistently evaluated in the exterior vicinity of the event horizon, $r \to h^+$, whereas, since the interior matter configuration is not of electromagnetic origin, the corresponding tensor vanishes in the interior limit, $F_a^{(-)} = 0$.

Although the hypersurface is null at the exact location of the event horizon, in the regions immediately outside, $r \to h^+$, the change in causal character allows one to define unit spacelike normal vectors. Accordingly, we define
$n_\mu^{(+)} = \frac{1}{\sqrt{f(r)}}\,\delta^r_\mu$ and $n^\alpha_{(+)} = \sqrt{f(r)}\,\delta^\alpha_r$.
Although the individual factors may diverge or vanish at $r = h$, in the limiting region $r \to h^+$ their combination remains well defined, yielding $n_\mu^{(+)} n^\mu_{(+)} = 1$. This corresponds to unit normalization in the immediate exterior region and reflects the change in causal character across the horizon. As we will see below in Eq.~\eqref{SaltoFaraday}, the quantity of interest is $\displaystyle \sigma \sim - \lim_{r \to h^+} F_{\alpha\beta}^{\text{(out)}}\, e^\alpha_a \, n^\beta_{(+)} u^a \sim - \lim_{r \to h^+} F_{t r}^{\text{(out)}}\, e^t_t \, n^r_{(+)} u^t$.
We note in the last expression that, in the limit $r \to h^+$, the non zero contribution of $n^\beta_{(+)} u^a$ remains finite. In particular, near the horizon one has $n^r_{(+)} \sim \sqrt{f(r)}$, while, as we will see below, $u^t \sim 1/\sqrt{f(r)}$, so that the limit of their product is well defined and is of order unity. Since the remaining factors in the above expression are regular in this limit, the full quantity is well defined and, as we will see below, yields a finite value for the effective surface density.

The tangent vectors to the hypersurface given by the event horizon, are defined as $e^\alpha_a = \frac{\partial x^\alpha}{\partial y^a}$, with $a = {t,\theta,\phi}$. In particular, $e^\alpha_a = 0$ for $\alpha = r$, and $e^\alpha_a = \delta^\alpha_a$ for $\alpha = {t,\theta,\phi}$. These tangent vectors satisfy the orthogonality condition $e^\alpha_a\, n_\alpha^{(+)} = 0$. The electric charge density $\sigma$ is obtained from the following equation, where the left-hand side is given by Eq \eqref{FaradayProyectado}:
\begin{equation} \label{SaltoFaraday}
[F_a]=F_a^{\,(+)}-F_a^{\,(-)}=F_a^{\,(+)} = -4\pi \sigma\, u_a \,,
\end{equation}
where $u_a$ is the projection of the exterior four-velocity associated with static observers in the limit $r \to h^+$. In particular,
$u_a = \displaystyle \lim_{r \to h^+}  \left(-\sqrt{f(r)}\,\delta^t_a\right)
$.
Thus, the factor $\sqrt{f(r)}$ appears on both sides of Eq.~\eqref{SaltoFaraday}. Even though $\sqrt{f(r)} \to 0$ as $r \to h^+$, solving for $\sigma$ leads to a ratio whose limit is well defined and remains finite, allowing this common factor to be consistently factored out. In this way, we obtain
\begin{equation} \label{SigmaEq}
\sigma = \lim_{r \to h^+} \frac{Q}{4\pi r^2} = \frac{Q}{4\pi h^2}.
\end{equation}

The quantity $\sigma$ can be interpreted as an effective surface charge density arising from the limiting behavior of the electromagnetic field near the event horizon. It is important to emphasize that, unlike in the standard thin-shell construction, where the energy–momentum sources responsible for the charge density are localized on the shell itself, no such localized sources are present in our case. Instead, the difference between the energy–momentum tensors in the interior and exterior regions gives rise to a non-trivial limiting behavior of the projected Maxwell tensor. In our setup, the effective charge density is localized in the exterior vicinity of the horizon, where the geometry changes smoothly through the continuity of the function $f(r)$ that determines the temporal and radial components of the metric. In this sense, $\sigma$ should be regarded as a quasi-local quantity defined through the limiting procedure $r \to h^+$, rather than as a fundamental surface source. Finally, the resulting expression for $\sigma$, Eq.~\eqref{SigmaEq}, provides a natural and finite characterization of this effective charge distribution, without introducing a genuine thin shell.

\section{Discussion and Summary}

In this work, we have investigated an alternative scenario for the internal structure of the Reissner–Nordström (RN) black hole, motivated by the simultaneous presence of a central non-integrable singularity and an inner horizon acting as a Cauchy horizon, which is commonly associated with instabilities and a potential loss of predictability in General Relativity. We have shown that the exterior RN geometry can instead be generated by an extended interior region containing an integrable singularity and lacking an inner horizon, thus replacing the conventional description based on a pointlike mass located at the origin.

We have analyzed the general conditions required to match an interior region with an exterior black hole geometry through the Israel–Darmois junction conditions \cite{Israel:1966rt}. We show that thermodynamic consistency across the interface requires the continuity of temperature, suggesting that the two regions must be in thermal equilibrium, while discontinuities in the tangential pressure can signal the presence of gravitational phase transitions. This thermodynamic interpretation provides an additional physical perspective on the matching between the interior and exterior geometries.

To explore the physical nature of the matter capable of sustaining such an interior geometry while generating a Reissner–Nordström exterior geometry, we considered two explicit realizations based on string distributions: a cloud of strings (CS) and a fluid of strings (FS). In the case of the cloud of strings, following the definition of conserved energy given in Ref. \cite{Aoki:2020prb}, the analysis of the conserved quantities associated with timelike Killing vectors reveals that the conserved energy diverges. Motivated by this feature, we introduce a new fluid-of-strings (FS) model in which the energy density can be interpreted as a geometrically screened version of the standard string-cloud distribution. This modification leads to configurations that, when extended to infinity, represent black holes with finite conserved energy, in contrast to the divergent conserved energy typically encountered in the string-cloud black hole.

In this way, for both interior regions constructed from the cloud-of-strings and the new fluid-of-strings distributions, we analyze the physical conditions that lead to the presence of an integrable singularity and the absence of an inner horizon. Furthermore, we determine the relations between the parameters of the two interior string distributions and the exterior Reissner–Nordström geometry for which the temperature remains continuous at the interface, thereby implying thermal equilibrium across the matching surface. In addition, we identify the regions of parameter space for which phase transitions at the event horizon may occur or be absent.

While configurations with an electric field present only in the exterior region are often described in terms of a charged thin shell \cite{Lemos:2021jtm,Abellan:2025bde}, our analysis carried out in Section \ref{seccionElectrico} indicates a different scenario. In the present case, the transition across the vicinity of the horizon is smooth, in the sense that the radial derivative of the function $f(r)$ is continuous, ensuring the continuity of the radial pressure. Moreover, the limiting behavior of the electromagnetic field near the horizon allows for the definition of an effective surface charge density $\sigma$. This quantity should be interpreted as a quasi-local object emerging from the limit $r \to h^+$, rather than as a fundamental surface source. The expression obtained for $\sigma$, Eq.~\eqref{SigmaEq}, provides a finite and natural characterization of this effective charge distribution.

It is worth mentioning that the framework developed in this work is not restricted to the Reissner–Nordström geometry. The general conditions obtained for interior geometries containing integrable singularities, together with their thermodynamic implications, may also be relevant for future studies of other black hole exterior spacetimes that exhibit a destructive central singularity and, in some cases, an inner horizon. In such situations, these exterior geometries could potentially be generated by an interior region endowed with an integrable singularity and lacking an inner horizon. Examples of black hole configurations that could be investigated within the framework presented here include black holes with cosmic void density profiles \cite{Lustosa:2025mxr}, black holes surrounded by dark matter \cite{Xu:2018wow}, hairy black holes \cite{Ovalle:2020kpd}, quintessential black holes \cite{Kiselev:2002dx,Estrada:2025sku} and black holes in theories beyond General Relativity \cite{Aros:2023tbh,Estrada:2024lhk,Estrada:2025ice,Estrada:2024uuu}.

It is also possible to consider, in future work, an extension of the formalism developed here to regular interior geometries, instead of those featuring an integrable singularity. In this context, Ref. \cite{Ovalle:2024wtv} proposes generic geometries as interiors of Schwarzschild black holes that exhibit an inner horizon. This motivates the study of the physical nature of alternative interior configurations for different exterior black hole geometries, as well as the implications associated with the presence or absence of such a horizon.
Among several potential examples, it is of interest to explore models that remain regular at Planckian scales \cite{Spallucci:2017aod}, as well as solutions based on nonlinear electrodynamics (NED) \cite{Ayon-Beato:1998hmi} In the specific case considered in this work, where the exterior geometry corresponds to Reissner–Nordström, it would be particularly relevant to investigate, as possible interior regions, regularized versions of this solution, for instance, those supported by effective NED sources (such as magnetic configurations) \cite{Bronnikov:2024izh} to analyze the conditions on the parameters required to satisfy the junction conditions.

\appendix

\section{A brief revision of the cloud of string model} \label{ApendiceNubeDeCuerdas}

In this appendix, we review the necessary tools for our study of the so-called cloud of strings from reference \cite{Letelier:1979ej}. A 1-brane string is parametrized by the parameters $\lambda^a = (\lambda^0, \lambda^1)$, which is embedded into spacetime as $x^\mu = x^\mu(\lambda)$. The trajectory of a 1-brane defines a two-dimensional worldsheet. The Nambu–Goto (NG) action for a 1-brane is proportional to the area of this worldsheet.
\begin{equation} \label{AccionNambuGoto}
    S_{NG}= \int \mathcal{M} \, dA = \int \mathcal{M} \sqrt{-h} d\lambda^0 d\lambda^0 \, \, ; \, \, h_{ab} = g_{\mu\nu} 
\frac{\partial x^\mu}{\partial \lambda^a} 
\frac{\partial x^\nu}{\partial \lambda^b} \,.
\end{equation}

Let $\mathcal{M}$ be a constant such that the action is dimensionless in natural units. Usually, $[\mathcal{M}] = [T_0/c] = [\text{Force}] = [\text{Energy}/L] = [L^{-2}]$, 
where $T_0$ and $c$ are the 1-brane tension and the speed of light, respectively. The induced metric is given by $h_{ab}$, while $\sqrt{-h}$ corresponds to its determinant. This reference introduces a bivector $\Sigma^{\mu\nu}$ that spans the two-dimensional time-like worldsheet of the string. It can be expressed as
\begin{equation}
\Sigma^{\mu\nu} = \epsilon^{AB} \frac{\partial x^\mu}{\partial \xi^A} \frac{\partial x^\nu}{\partial \xi^B},
\end{equation}
where $\epsilon^{AB}$ is the two-dimensional Levi-Civita symbol with components $\epsilon^{01} = -\epsilon^{10} = -1$. The action in \eqref{AccionNambuGoto} leads to an energy-momentum tensor for a 1-brane string. This reference shows that for an ensemble or cloud of strings, this tensor takes the form:
\begin{equation} \label{TensorEMNubeCuerdas}
    T^{\mu\nu} = - \frac{\rho_{cs} \, \Sigma^{\mu\alpha} \Sigma_{\alpha}^{\;\;\nu}}{\sqrt{-h}} \,\, ;\,\, h = \frac{1}{2} \Sigma_{\mu\nu} \Sigma^{\mu\nu} .
\end{equation}

For the energy–momentum tensor reported in this reference, Eq.~\eqref{TensorEMNubeCuerdas}, the only nonzero components are $T^0_0 = T^1_1$. The conservation equation $\nabla_\mu T^{\mu\nu} = 0$ implies that the structure of this tensor is given by
\begin{equation} \label{TensorEMNubeCuerdas1}
-T^0_0 = - T^1_1 = \rho = \sqrt{-h}\,\rho_{cs} = \frac{a}{r^2},
\end{equation}
where $\rho_{cs}$ can be interpreted as the proper rest energy density of the string cloud. 

\section{Some general aspects of the fluid of strings model} \label{ApendiceFluidoDeCuerdas}
As previously discussed, the original cloud of strings framework was subsequently extended in Ref. \cite{Letelier:1983du} in order to incorporate the presence of an effective pressure. Within this generalized description, the energy--momentum tensor takes the form
\begin{equation} \label{TensorEMfluid}
    T^{\mu\nu}
= \left(p + \rho_{cs} \sqrt{-h}\right)
\frac{\Sigma^{\mu\lambda}\Sigma^{\nu}{}_{\lambda}}{(-h)}
+ p\, g^{\mu\nu},
\end{equation}
where $p$ and $\rho_{cs}$ denote, respectively, the pressure and the energy density associated with the string fluid. Considering a line element of the form \eqref{ElementoDeLineaInterior}. As indicated in Ref. \cite{Soleng:1993yr}, due to the symmetries of the metric, the only nonvanishing components of $\Sigma^{\mu\lambda}$ are $\Sigma^{tr}$ and $\Sigma^{\theta\phi}$; therefore, the determinant of the induced metric satisfies $h<0$.

\bibliography{mybib.bib}

\end{document}